\documentclass[aps,prl,twocolumn,showpacs,preprintnumbers,amsmath,amssymb,superscriptaddress]{revtex4}%

\usepackage{graphicx}%
\usepackage{dcolumn}
\usepackage{amsmath}
\usepackage{color}

\begin{document}

\title{Evolution of two-gap behavior of the superconductor FeSe$_{1-x}$ }
\author{R.~Khasanov}
 \email[Corresponding author: ]{rustem.khasanov@psi.ch}
 \affiliation{Laboratory for Muon Spin Spectroscopy, Paul Scherrer
Institute, CH-5232 Villigen PSI, Switzerland}
\author{M.~Bendele}
 \affiliation{Laboratory for Muon Spin Spectroscopy, Paul Scherrer
Institute, CH-5232 Villigen PSI, Switzerland}
 \affiliation{Physik-Institut der Universit\"{a}t Z\"{u}rich,
Winterthurerstrasse 190, CH-8057 Z\"urich, Switzerland}
\author{A.~Amato}
 \affiliation{Laboratory for Muon Spin Spectroscopy, Paul Scherrer
Institute, CH-5232 Villigen PSI, Switzerland}
\author{K.~Conder}
 \affiliation{Laboratory for Developments and Methods, Paul Scherrer Institute,
CH-5232 Villigen PSI, Switzerland}
\author{H.~Keller}
 \affiliation{Physik-Institut der Universit\"{a}t Z\"{u}rich,
Winterthurerstrasse 190, CH-8057 Z\"urich, Switzerland}
\author{H.-H.~Klauss}
 \affiliation{IFP, TU Dresden, D-01069 Dresden, Germany}
\author{H.~Luetkens}
 \affiliation{Laboratory for Muon Spin Spectroscopy, Paul Scherrer
Institute, CH-5232 Villigen PSI, Switzerland}
\author{E.~Pomjakushina}
 \affiliation{Laboratory for Developments and Methods, Paul Scherrer Institute,
CH-5232 Villigen PSI, Switzerland}

\begin{abstract}
The superfluid density, $\rho_s$, of the iron chalcogenide
superconductor, FeSe$_{1-x}$, was studied as a function of
pressure by means of muon-spin rotation. The zero-temperature
value of $\rho_s$ increases with increasing transition temperature
$T_c$ (increasing pressure) following the tendency observed
for various Fe-based and cuprate superconductors. The analysis of
$\rho_s(T)$ within the two-gap scheme reveals that the effect
on both, $T_c$ and $\rho_s(0)$, is entirely determined by the
band(s) where the large superconducting gap develops, while the
band(s) with the small gap become practically unaffected.
\end{abstract}

\maketitle

Since the discovery of Fe-based high-temperature superconductors
(HTS) much effort is devoted to the investigation of their
superconducting mechanism.  While some properties of Fe-based HTS
are reminiscent of the cuprate HTS (as, {\it e.g.}, their layered
structure, the proximity to a magnetic phase, the universal
``Uemura'' scaling between the superfluid density, $\rho_s$, and
the transition temperature, $T_c$), the differences between both
compounds families are much more remarkable. Hence, the
superconductivity in Fe-based HTS originates within the
$d$-orbitals of the Fe ion, which are normally expected to lead to
pair-breaking effects \cite{Cao08Singh08}. For the Fe-based HTS,
several disconnected Fermi-surface sheets contribute to 
superconductivity, as revealed by angle-resolved photoemission
spectroscopy \cite{Ding08Zhao08Zabolotnyy08,Kondo08}. Furthermore,
indications for multi-gap superconductivity was obtained from
tunneling \cite{Gonelli09,Samuely09}, magnetic torque
\cite{Weyeneth09_1Weyeneth09_2}, point contact \cite{Szabo08} and
infrared spectroscopy \cite{Li09} experiments, as well as from
specific heat \cite{Mu09}, first and second critical field
\cite{Ren08_2,Hunte08}, and superfluid density
\cite{Malone08,Hiraishi08,Khasanov09_BKFAKhasanov09_Sr122,Khasanov08_FeSe}
studies.
The multi-gap superconducting state positions the Fe-based HTS together
with MgB$_2$ -- the most famous double-gap superconductor discovered to date.
However, it is worth mentionning that in spite of the fact that
the two-gap superconductivity was detected for Fe-based HTS
belonging to different families (as {\it e.g.} 1111:
\cite{Gonelli09,Hunte08,Malone08,Weyeneth09_1Weyeneth09_2}; 122:
\cite{Ding08Zhao08Zabolotnyy08,Samuely09,Mu09,
Ren08_2,Szabo08,Hiraishi08, Khasanov09_BKFAKhasanov09_Sr122};
011: \cite{Khasanov08_FeSe}) a systematic study of this
phenomenon within one given family was not yet performed.
In this paper we report on the evolution of two-gap behavior
in the iron chalcogenide superconductor FeSe$_{1-x}$. The transition
temperature was changed within the range $8.3\lesssim
T_c\lesssim12.8$~K by applying an external pressure $p$ between 0 and
0.84~GPa.
At each particular pressure the superfluid density $\rho_s$ was
obtained from the in-plane magnetic penetration depth
$\lambda^{-2}_{ab}(T)\propto\rho_s$ studied by means of muon-spin
rotation, $\mu$SR.
The analysis of $\lambda_{ab}^{-2}(T,p)$, performed by
solving self consistently the gap equations derived within the
two-gap scheme \cite{Bussmann-Holder04,Bussmann-Holder09}, reveals
that the main effect on $T_c(p)$ and
$\lambda_{ab}^{-2}(T,p)\propto \rho_s(T,p)$ arises from the energy
band(s) where the large superconducting gap, $\Delta_1$, develops.
The zero-temperature values of $\Delta_{1}$, the contribution of
this gap to the superfluid density $\lambda_{ab,1}^{-2}$, as well
as the effective coupling constant $\Lambda_{11}$ increase almost
linearly with increasing $T_c$ (increasing pressure). In contrast,
the contribution of the small gap and thus $\Delta_2$,
$\lambda_{ab,2}^{-2}$, and $\Lambda_{22}$, is practically pressure
independent. Our results imply, therefore, that the transition
temperature in FeSe$_{1-x}$ is entirely determined by the
intraband interaction within the band(s) where the dominant gap is
opened.

The sample with the nominal composition FeSe$_{0.94}$ was
prepared by solid state reaction similar to that described in
Refs.~\onlinecite{Hsu08,Margadonna09,Pomjakushina09}. Powders of
minimum purity 99.99\% were mixed in appropriate ratios, pressed
and sealed in a double-walled quartz ampoule. The sample was
heated up to 700$^{\rm o}$C followed by annealing at 400$^{\rm
o}$C  \cite{Pomjakushina09}.
The pressure was generated in a CuBe  piston-cylinder type of cell
especially designed to perform $\mu$SR experiments under pressure
\cite{Andreica01}. As a pressure transmitting medium 7373 Daphne
oil was used. The pressure was measured in situ by monitoring the
pressure shift of the superconducting transition temperature of Pb
and/or In.
The $\mu$SR experiments were carried out at the $\mu$E1 beam line,
Paul Scherrer Institute, Switzerland. Zero-field (ZF) and
transverse-field (TF) $\mu$SR experiments were performed at
temperatures ranging from 0.24 to 50~K. For TF measurements the
external magnetic field $\mu_0H=10$~mT was applied perpendicular
to the muon-spin polarization. Typical counting statistics was
$\sim5-7\cdot 10^{6}$ positron events for each data point.

The results of the ZF $\mu$SR experiments were previously reported
in Ref.~\onlinecite{Bendele09}. It was shown that up to $p\simeq
0.8$~GPa the ZF response of FeSe$_{0.94}$ is determined by the
contribution of the dilute Fe moments, in analogy with what was
observed for ambient pressure measurements of FeSe$_{0.85}$
\cite{Khasanov08_FeSe}. At $p=0.84$~GPa static magnetism was found
to occupy approximately 10\% of the sample volume at $T\simeq T_c$
and it decreases down to $\sim5$\% at $T\simeq 0.25$~K
\cite{Bendele09}.

\begin{figure}[t]
\includegraphics[width=0.7\linewidth]{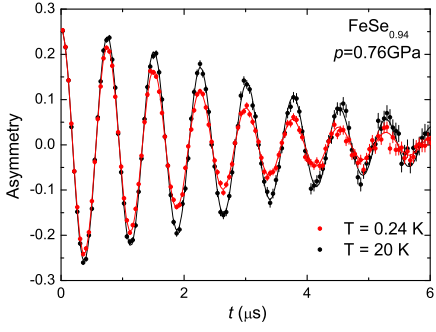}
 \vspace{-.3cm}
\caption{ (Color online) TF $\mu$SR time-spectra ($\mu_0H=10$~mT) of
FeSe$_{0.94}$ measured below ($T=0.24$~K) and above ($T=20$~K) the
superconducting transition temperature ($T_c\simeq13$~K) at
$p=0.76$~GPa. The stronger damping in the superconducting state is
due to the formation of the vortex lattice.}
 \label{fig:TF-signal}
\end{figure}

Figure~\ref{fig:TF-signal} shows the TF $\mu$SR time-spectra
measured at $p=0.76$~GPa above ($T=20$~K) and below ($T=0.24$~K)
the superconducting transition temperature ($T_c\simeq13$~K). The
stronger relaxation of the muon-spin polarization at 0.24~K
relative to 20~K is due to the formation of the vortex lattice at
$T<T_c$.
The TF $\mu$SR data were analyzed by using the
functional form:
\begin{eqnarray}
A(t)&=&A_S(t)+A_{PS}(t)  \nonumber \\
&=& A_{S,0}\;e^{-\Lambda t}\;e^{-\sigma_S^2t^2/2}\cos(\gamma_\mu
B_{S}\;t+\phi) \nonumber \\
&&+A_{PS,0}\;e^{-\sigma_{PS}^2\ t^2/2}\cos(\gamma_\mu
B_{PS}\;t+\phi).
 \label{eq:TF}
\end{eqnarray}
Here the indexes $S$ and $PS$ denote the sample and the pressure
cell, respectively, $A_0$ is the initial asymmetry, $\Lambda$ is
the exponential relaxation rate caused by the presence of diluted
Fe moments \cite{Khasanov08_FeSe},
$\gamma_\mu=2\pi\cdot135.5$~MHz/T is the muon gyromagnetic ratio,
$B$ is the internal field, and $\phi$ is the initial phase of the
muon-spin ensemble. The Gaussian relaxation rate, $\sigma_{PS}$,
reflects the depolarization due to the nuclear magnetism of the
pressure cell, while $\sigma_{S}$ represents the depolarization in
the sample arising from the nuclear moments and from the vortex
lattice (see below).
Each set of TF $\mu$SR data taken at constant pressure was fitted
simultaneously with $A_{S,0}$, $A_{PS,0}$, $B_{PS}$,
$\sigma_{PS}$, $\Lambda$, and $\phi$, as common parameters, and
$B_{S}$ and $\sigma_S$  as individual parameters for each
temperature point. The exponential relaxation rate $\Lambda$ was
assumed to be temperature independent in accordance with the
results of ZF $\mu$SR experiments \cite{Bendele09}.

\begin{figure}[t]
\includegraphics[width=1.0\linewidth]{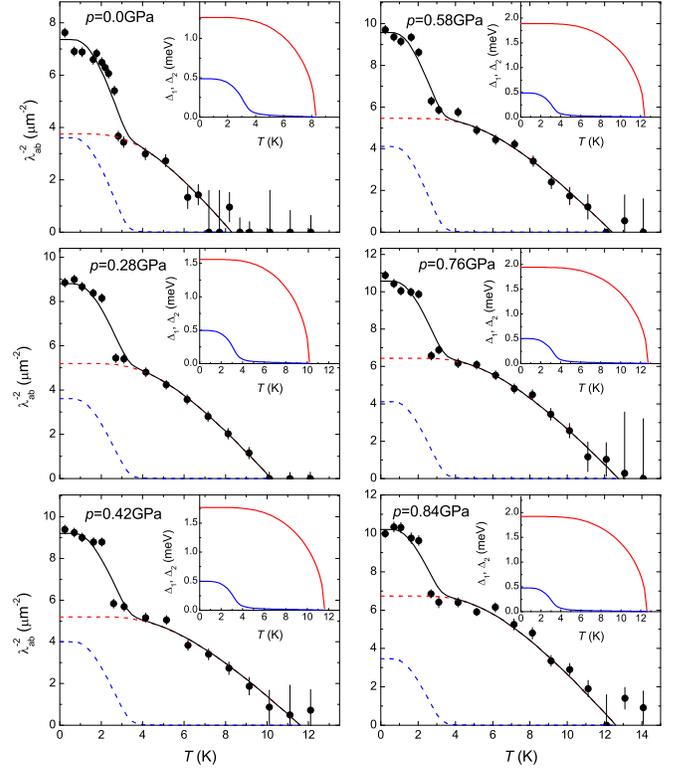}
 \vspace{-.7cm}
\caption{(Color online)  Temperature dependence of
$\lambda^{-2}_{ab}\propto \rho_s$ of FeSe$_{0.94}$ measured at
$p=0.0$, 0.28, 0.42, 0.58, 0.76, and 0.84~GPa. The solid and the
dashed lines are the theoretical curves obtained within the
framework of the two-gap model described in the text. The insets
show the temperature dependences of the large ($\Delta_1$) and the
small ($\Delta_2$) gap. }
 \label{fig:superfluid_common}
\end{figure}

In an anisotropic powder sample the magnetic penetration depth
$\lambda$ can be extracted from the Gaussian relaxation rate
$\sigma_{sc}(T)=[\sigma_S^2(T)-\sigma_{nm}^2]^{1/2} \propto
1/\lambda^{2}(T)$, which probes the second moment of the magnetic
field distribution in a superconductor in the mixed state
\cite{Khasanov08_FeSe,Khasanov09_SmNd1111,Brandt88}. Here
$\sigma_{nm}$ is the nuclear moment contribution measured at
$T>T_c$. $\sigma_{sc}$ can be converted into $\lambda_{ab}$ via
\cite{Khasanov08_FeSe,Khasanov09_SmNd1111}:
\begin{equation}
\sigma_{sc}^2/\gamma_\mu^2=0.00126\;\Phi_0^2/\lambda_{ab}^{\ 4},
 \label{eq:lambda_ab}
\end{equation}
where $\Phi_0=2.068\cdot10^{-15}$~Wb is the magnetic flux quantum.
The measured $\lambda^{-2}_{ab}(T,p)$ of FeSe$_{0.94}$ at
$p=0.0$, 0.28, 0.42, 0.58, 0.76, and 0.84~GPa are shown in
Fig.~\ref{fig:superfluid_common}.

The experimental $\lambda^{-2}_{ab}(T)$ data were analyzed by
using the two-gap model presented recently in Refs.~\onlinecite{Bussmann-Holder04,Bussmann-Holder09}.
Following \cite{Bussmann-Holder04}, the coupled gap equations for
a superconductor with $k$-dependent energy gaps: $\Delta_1$,
$\Delta_2$, intraband pairing potentials: $V_{11}$, $V_{22}$, and
interband interaction potentials: $V_{12}$, $V_{21}$ are
determined as:
\begin{eqnarray}
\Delta_i(k_i)&=&\sum_{j=1,2}\sum_{k_j'} \frac{V_{i,j}(k_i,k_j')\, \Delta_j(k_j')}{2\ \sqrt{E^2(k_j')+\Delta_j^2(k_j')}} \nonumber\\
&& \times \tanh \frac{\sqrt{E^2(k_j')+\Delta_j^2(k_j')}}{2k_BT}
 \label{eq:gap-standart}
\end{eqnarray}
Here $i=1,2$ is the band index and the sums are taken within the
corresponding energy bands in  $k$-space. Considering
isotropic $s$-wave gaps [$\Delta_i(k_i)=\Delta_i$] and neglecting
the momentum dependence of the band energies,
Eq.~(\ref{eq:gap-standart}) reduces to:
\begin{equation}
\Delta_i=\sum_{j=1,2}\ \int_0^{\omega_{D_i}} \frac{N_j(0)\,
V_{i,j}\, \Delta_j }{\sqrt{E^2+\Delta_{j}^2}}
\tanh\frac{\sqrt{E^2+\Delta^2_{j}}}{2k_BT} dE,
 \label{eq:gap-long}
\end{equation}
Here $\omega_{D_i}$ is the phonon cutoff frequency (Debye
frequency; note that these cutoffs are expected to be different
for both bands $\omega_{D_1}\neq\omega_{D_2}$
\cite{Bussmann-Holder09}) and $N_{i}(0)$ is the partial density of
states at the Fermi level. For convenience the sums were also
converted into integrals \cite{Bussmann-Holder09}.
A further simplification of Eq.~(\ref{eq:gap-long}) can be
made by using the notation of the coupling constant
$\Lambda_{ij}=N_i(0)V_{ij}$  introduced by Kogan {\it et al.}
\cite{Kogan09} and assuming similar cutoff frequencies for both
bands ($\omega_{D_1}=\omega_{D_2}=\omega_D$):
\begin{equation}
\Delta_i=\sum_{j=1,2}\ \Lambda_{ij}\Delta_j\int_0^{\omega_D}
\frac{1}{\sqrt{E^2+\Delta_{j}^2}}
\tanh\frac{\sqrt{E^2+\Delta_{j}^2}}{2k_BT} dE.
 \label{eq:gap-short}
\end{equation}
The advantages to use the above equation rather than
Eq.~(\ref{eq:gap-long}) is that (i) within the notation of Kogan
{\it et al.} \cite{Kogan09} $\lambda_{12}=\lambda_{21}$ and (ii)
the total number of parameters needed to evaluate $\Delta_1(T)$ and
$\Delta_2(T)$ reduces from 8 in case of
Eq.~(\ref{eq:gap-long}) to 4 in case of
Eq.~(\ref{eq:gap-short}).

With the known $\Delta_1(T)$ and $\Delta_2(T)$, $\lambda_{ab}^{-2}$ can be 
obtained by decomposing it into two components
$\lambda_{ab,1}^{-2}$ and $\lambda_{ab,2}^{-2}$ so that:
\begin{equation}
\lambda_{ab}^{-2}(T)=\lambda_{ab,1}^{-2}(T)
+\lambda_{ab,2}^{-2}(T)
\end{equation}
with \cite{Tinkham75}:
\begin{equation}
\frac{\lambda_{ab,i}^{-2}(T)}{\lambda_{ab,i}^{-2}(0)}=  1+
2\int_{\Delta_i(T)}^{\infty}\frac{\partial f}{\partial
E}\frac{E}{\sqrt{E^2-\Delta_i(T)^2}}\  dE \  . \nonumber
\end{equation}
%
%
Here $f=[1+\exp(E/k_BT)]^{-1}$ is  the Fermi function.

%
\begin{figure}[htb]
\includegraphics[width=1.0\linewidth]{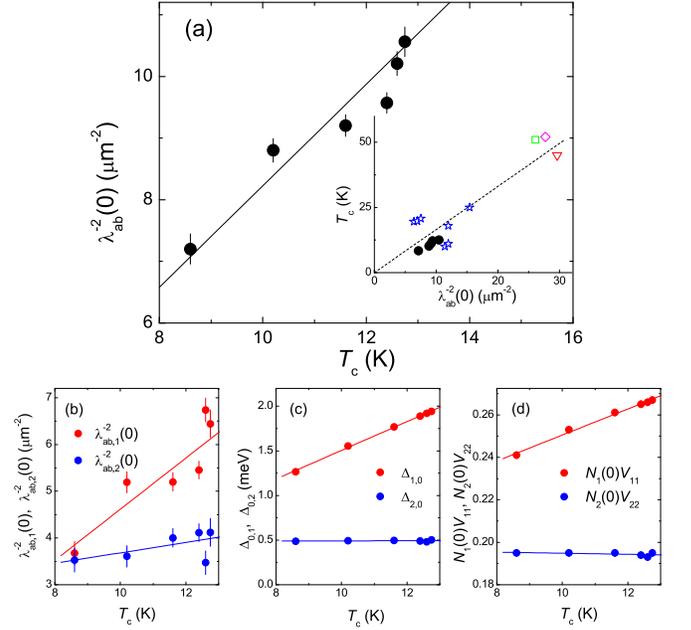}
 \vspace{-0.7cm}
\caption{(Color online) (a) Dependence of $\lambda^{-2}_{ab}(T=0)$ on $T_c$. The inset is the ``Uemura relation'' for Fe-based HTS with some of data obtained to date (see Ref.~\onlinecite{Khasanov08_FeSe,Khasanov09_BKFAKhasanov09_Sr122, Khasanov09_SmNd1111,Luetkens08Drew08Amato09Carlo09Goko09}).   (b), (c), and (d) Dependence of the parameters obtained in analysis of $\lambda_{ab}^{-2}(T,p)$ within the framework of two-gap model (see text for details).}
 \label{fig:parameters}
\end{figure}

\begin{table*}[htb]
\caption[~]{\label{Table:results} Summary of the pressure studies
of FeSe$_{1-x}$. The meaning of the parameters is: $p$ -- pressure;
$T_c$ -- transition temperature; $\omega_D$ -- Debye frequency;
$\Lambda_{12}$ -- interband coupling constant,
$\Lambda_{11}/\Lambda_{22}$, $\Delta_1(0)/\Delta_2(0)$,
$\lambda_{ab,1}^{-2}(0)/\lambda_{ab,2}^{-2}(0)$ -- intraband
coupling constant, zero-temperature value of the gap, zero
temperature value of superfluid density component within the band
1/2, respectively.}
\begin{center}
 \vspace{-0.5cm}
\begin{tabular}{ccclllcccccc}\\
 \hline
 \hline
$p$&$T_c$&$\omega_D$&$\Lambda_{11}$& $\Lambda_{22}$&$\Lambda_{12}$&$\Delta_1(0)$&$\Delta_2(0)$&  $\lambda_{ab,1}^{-2}(0)$&$\lambda_{ab,2}^{-2}(0)$\\
(GPa) &(K)&(meV)&&&&(meV)&(meV)&($\mu$m$^{-2}$)&($\mu$m$^{-2}$)\\
\hline
0.00 &8.3(1) &40.0  &0.241(1)&0.195(1) &0.0005 &1.27(1)&0.487(6)&3.67(25)&3.53(25)\\
0.28 &10.2(1)&40.36 &0.253(1)&0.195(1) &0.0005 &1.56(1)&0.494(6)&5.19(23)&3.61(23)\\
0.42 &11.6(1)&40.54 &0.261(1)&0.195(1) &0.0005 &1.77(1)&0.498(6)&5.20(20)&4.00(18)\\
0.58 &12.4(1)&40.74 &0.265(1)&0.194(1) &0.0005 &1.89(1)&0.491(6)&5.45(19)&4.11(20)\\
0.76 &12.8(1)&40.98 &0.267(1)&0.195(1) &0.0005 &1.94(1)&0.504(6)&6.44(29)&4.12(25)\\
0.84 &12.6(1)&41.08 &0.266(1)&0.193(1) &0.0005 &1.92(1)&0.480(6)&6.74(29)&3.47(29)\\
 \hline \hline \\
\end{tabular}
   \end{center}
\end{table*}

The analysis of $\lambda_{ab}^{-2}(T,p)$ by using the above described
model  was made by solely evaluating $\Lambda_{22}$,
$\lambda_{ab,1}^{-2}(0)$, and $\lambda_{ab,2}^{-2}(0)$. The
parameters $\Lambda_{11}$, $\Lambda_{12}$, and
$\omega_D$ were taken as follows: \\
$\Lambda_{12}$: Our numerical analysis reveals that the step-like
change of $\lambda_{ab}^{-2}(T)$ at $T\simeq 2.5$~K (see
Fig.~\ref{fig:superfluid_common}) requires the interband coupling
constant $\Lambda_{12}$ to be very small ($\Lambda_{12}\sim
10^{-3}$ or smaller). This implies that the band(s), where the large
and the small superconducting energy gaps are open, become only
weakly coupled. Note that a similarly small interband coupling constant was
obtained by Kogan {\it et al.} for the
superconductor V$_3$Si \cite{Kogan09}.\\
$\Lambda_{11}$: The fact that the interband coupling in
FeSe$_{1-x}$ is weak thus suggests that the transition temperature
$T_c$ is mainly determined by the coupling within the band(s) where
the large superconducting gap is opened. Assuming that the larger
gap is $\Delta_1$, $T_c$ is defined when $\Delta_1(T)=0$, so that
according to Eq.~(\ref{eq:gap-short}):
\begin{equation}
\Lambda_{11}\simeq\left[\int_0^{\omega_D}
\frac{dE}{E}\tanh\frac{E}{2k_BT_c} \right]^{-1}.
 \label{eq:Tc}
\end{equation}
With $T_c(p)$ measured independently (see Ref.~\cite{Bendele09})
Eq.~(\ref{eq:Tc}) allows one to obtain the value of the intraband
coupling constant $\Lambda_{11}$ for each particular pressure.\\
$\omega_D$: The ambient pressure value of the cutoff phonon
frequency (Debye frequency) $\omega_D(p=0)\simeq40$~meV was taken
from Ref.~\cite{Phelan09}. The increase of $\omega_D$ with
increasing pressure was assumed to follow:
\begin{equation}
\omega_D(p)=\omega_D(0)(1+\gamma p/B),
 \label{eq:omega_D}
\end{equation}
which is the consequence of the Gr\"uneisen equation
$\gamma=-d\ln\omega_D/d\ln V$ ($\gamma$ is the Gr\"uneisen
parameter, $B\simeq31$~GPa is the bulk modulus
\cite{Margadonna09}, and $V$ is the sample volume). The Gr\"uneisen
parameter was assumed to be $\gamma\approx1$ in analogy with
Ref.~\onlinecite{Huang09}. We should also emphasize that the
parameters of the above described model are not very sensitive to
the exact value of $\omega_D$. As an example, the increase
(decrease) of $\omega_D$ by a factor of 2 leads to a corresponding
decrease(increase) of $\Lambda_{11}$ obtained from
Eq.~(\ref{eq:omega_D}) by $\simeq15(18)$\%. This make our
assumption about using a similar cutoff phonon frequency for both
bands [$\omega_{D_1}=\omega_{D_2}$, see Eq.~(\ref{eq:gap-short})]
to be rather reliable. Note that similar conclusion was also
reached by Kogan {\it et al.} \cite{Kogan09_Reply}.

The parameters obtained from the analysis of
$\lambda_{ab}^{-2}(T)$ by means of the model described above are
summarized in Table~\ref{Table:results}. The value of the
interband coupling constant $\Lambda_{12}=0.005$ was kept fixed.
The red and the blue dashed lines in
Fig.~\ref{fig:superfluid_common} correspond to the contribution of
the large, $\lambda_{ab,1}^{-2}$, and the small,
$\lambda_{ab,2}^{-2}$, superconducting gaps to the total
superfluid density, solid lines. The temperature dependences of
the large, $\Delta_1$, and the small, $\Delta_2$, gaps are shown
in the corresponding insets.

In order to check how the change of $T_c$ affects the energy bands
where the large and the small superconducting gaps are supposed to
be open, we plot in Fig.~\ref{fig:parameters} the parameters
$\lambda_{ab}^{-2}(0)$, $\lambda_{ab,1}^{-2}(0)$,
$\lambda_{ab,2}^{-2}(0)$, $\Delta_1(0)$, $\Delta_2(0)$,
$\Lambda_{11}=N_1(0)\;V_{11}$, and $\Lambda_{22}=N_2(0)\;V_{22}$
as a function of the transition temperature $T_c$.
From the obtained data the following conclusions can be
drawn:
(i) The zero-temperature superfluid density
$\rho_s(0)\propto\lambda_{ab}^{-2}(0)$ increases with increasing
$T_c$ thus following the ``Uemura'' relation established recently
for various Fe-based HTS, see Fig.~\ref{fig:parameters}~(a) and
Refs.~\onlinecite{Khasanov08_FeSe,Khasanov09_BKFAKhasanov09_Sr122,
Khasanov09_SmNd1111,Luetkens08Drew08Amato09Carlo09Goko09}.
(ii) The electronic bands, where the large and the small gap are
opened, are affected by the pressure quite differently. The increase
of $T_c$ with pressure leads to an almost linear increase of the
superfluid density component $\lambda_{ab,1}^{-2}(0)$, the
superconducting energy gap $\Delta_{1,0}$ as well as the effective
coupling constant $\Lambda_{11}=N_1(0)\; V_{11}$. On the other
hand, both $\Delta_{2,0}$ and $\Lambda_{22}=N_2(0)\; V_{22}$ stay
almost constant, while $\lambda_{ab,2}^{-2}(0)$ increases with
increasing $T_c$ only slightly [see
Figs.~\ref{fig:parameters}~(b), (c) and (e)]. Bearing in mind that
the ``large gap'' and the ``small gap'' bands are only weakly
coupled (the interband coupling constant $\Lambda_{12}$ is
estimated to be of the order of $5\cdot10^{-4}$ or less, see
Table~\ref{Table:results}) one may conclude that in the range of
$0\leq p\leq 0.84$~GPa the pressure effect on both $T_c$ and
$\lambda_{ab}^{-2}$ is solely determined by the bands exhibiting
the large superconducting gap.

To conclude, the superfluid density
$\rho_s\propto\lambda_{ab}^{-2}$ was studied as a function of
pressure and temperature in the superconductor FeSe$_{1-x}$ by means
of $\mu$SR. The analysis of $\rho_s(T)$ within a
two-gap scheme reveals that the effect on both, $T_c$ and
$\rho_s(0)$, is entirely determined by the band(s) where the large
superconducting gap develops. Our results suggests that for 011
family of Fe-based HTS the intraband interaction is most probably
the leading pairing interaction determining the superconducting
properties.

This work was performed at the S$\mu$S Paul
Scherrer Institute, Switzerland.  The work of MB was supported by
the Swiss National Science Foundation. The work of EP was
supported by the NCCR program MaNEP.

\end{document}